\documentclass[pra,10pt,aps,twocolumn,amsmath,amssymb,altaffilletter]{revtex4-1}
\usepackage[utf8]{inputenc}
\usepackage     [T1]                    {fontenc}
\usepackage                             {color}
\usepackage                             {graphicx}
\usepackage     [english]               {babel}
\usepackage     [colorlinks,citecolor=black,linkcolor=black,urlcolor=blue,bookmarks=false,hypertexnames=true]           {hyperref}
\usepackage     [usenames,dvipsnames]   {pstricks}
\usepackage                             {bm}
\usepackage                             {textcomp}
\usepackage				{isotope}
\usepackage				{bigints}

\begin{document}

\title{Thermalization in Solid-State NMR Controlled by Quantum Chaos in Spin Bath}

\author{Walter Hahn, V. V. Dobrovitski}
% \email{w.hahn@tudelft.nl}
% \email{v.v.dobrovitski@tudelft.nl}
\affiliation{QuTech, Delft University of Technology, Lorentzweg 1, 2628 CJ Delft, The Netherlands}

% \date{\today}

\begin{abstract}
We theoretically investigate thermalization and spin diffusion driven by a quantum spin bath for a realistic solid-state NMR experiment. We consider polycrystalline L-alanine, and investigate how the spin polarization spreads among several $\isotope[13]{C}$ nuclear spins, which interact via dipole-dipole coupling with the bath of strongly dipolar-coupled $^1$H nuclear (proton) spins. We do this by using direct numerical simulation of the many-spin time-dependent Schr\"odinger equation. We find that, although the proton spins located near the carbon sites interact most strongly with the $^{13}$C spins, this interaction alone is not enough to drive spin diffusion and thermalize the $^{13}$C nuclear spins. We demonstrate that the thermalization within the $^{13}$C subsystem is driven by the collective many-body dynamics of the proton spin bath, and specifically, that the onset of thermalization among the $^{13}$C spins is directly related to the onset of chaotic behavior in the proton spin bath. Therefore, thermalization and spin diffusion within the $^{13}$C subsystem is controlled by the proton spins located far from the C sites. In spite of their weak coupling to the $^{13}$C spins, these far-away protons help produce a network of strongly coupled proton spins with collective dynamics, that drives thermalization. 
\end{abstract}

\maketitle

\section{Introduction}
Thermalization via spin diffusion is ubiquitous in many-spin systems, governing their most fundamental dynamical properties~\cite{bloembergen,bloembergen2,erg}. Understanding the dynamics of thermalization and spin diffusion is crucial for a broad range of applications, from control and protection of the qubit dynamics in the quantum information processing devices~\cite{spintronics,chumak,zoller,budker}, to elucidating the structure and the functions of the biomolecules via solid-state NMR (ssNMR)~\cite{kern,ladizhansky,oschkinat,hong,hong2,grommek,Wittmann2016}. However, thermalization dynamics in quantum systems is not fully understood due to complexity of the problem. Thermalization is a result of the quantum non-equilibrium evolution of a large number of interacting spins. Dynamics of the spin diffusion and its quenching by disorder is tightly related to important but poorly studied phenomena, such as onset of quantum chaos~\cite{pastawski,absence_fine,santos_vector,cazalilla} in many-spin systems and in few-spin systems coupled to their many-spin environments.

Progress in understanding thermalization and spin diffusion has been achieved recently by studying well-controlled systems, such as chains of trapped ions~\cite{blatt_2,blatt} and atomic 1D and 2D lattices~\cite{Parsons,bloch,browaeys}. At the same time, ssNMR constitutes an important test case for thermalization and spin diffusion in quantum many-spin systems~\cite{NEVZOROV,haase,Alvarez,waug}; these phenomena underlie many fundamental ssNMR techniques such as coherence and polarization transfer, measurement of the inter-nuclear distances via spin diffusion rates, and ssNMR-based structural analysis~\cite{hong,oschkinat,KONEKE2010197}. However, the typical systems studied in real ssNMR experiments are very complex, comprising many different nuclear species having anisotropic chemical shifts, subject to time-periodic modulation of the Hamiltonian parameters via magic angle spinning (MAS), and controlled by strong external driving. Microscopic description of thermalization in such systems is a challenge: it is still not clear which microscopic effects govern thermalization and polarization/coherence transfer, and to which extent the existing theoretical tools can describe spin diffusion at the microscopic level in systems of such complexity~\cite{santos_vector,fine_phase,olshanii,dykman,abanin}.

In this article, we theoretically study thermalization and spin diffusion in realistic ssNMR settings, considering polycrystalline powder of alanine, and taking into account all relevant experimental details (spatial arrangement of the nuclei, their anisotropic chemical shifts, periodic modulation of the Hamiltonian by MAS, etc). We argue that the dynamics of thermalization in ssNMR experiments (in our case, thermalization between different $^{13}$C spins) is governed by collective many-body quantum effects, namely by emergence of the chaotic dynamics in the surrounding spin environment (in our case, the nuclear spins of the hydrogen atoms, i.e.\ the proton spins). We show that even those proton spins that are located far away from the $^{13}$C spins, and thus are practically decoupled from them, still critically affect thermalization, due to essential many-body nature of the collective chaotic dynamics of the proton spin environment. Traditionally, spin diffusion in ssNMR experiments is described via semi-phenomenological Bloch-Redfield-type theories \cite{meier,slichter,abragam}; their validity relies on the heuristic notion of the "network of strongly coupled proton spins". Our work provides formalization for this notion in terms of the spectral properties of the proton spin subsystem. 

\section{Qualitative considerations}

Thermalization in many-spin systems usually occurs via flip-flop transitions, when two coupled spins exchange their polarization. If the local magnetic fields at the two spin sites differ significantly (in our case, different $^{13}$C spins have different Larmor frequencies due to different chemical shifts) the flip-flop transitions are suppressed because of the mismatch in the Zeeman energies, and the spin diffusion may be quenched (onset of localization)~\cite{feher,anderson}. In such case, the external time-dependent noise can assist spin diffusion: the time-varying random local fields occasionally bring the Larmor frequencies of the spins close to each other, enabling the flip-flops, and thus promoting thermalization. Specifically, the $^{13}$C spins occupying two chemically inequivalent sites with different chemical shifts, can equalize their polarizations if assisted by the noise created by the surrounding bath of proton spins~\cite{meier,veshtort,C1CP22662B,C1CP00004G} (proton-driven spin diffusion, PDSD). 

Viewed in this way, thermalization via proton-driven spin diffusion is an ssNMR representative of a typical situation of a few-spin central system ($^{13}$C spins) coupled to the large quantum spin environment ($^1$H spins). 

The standard theoretical analysis of the proton-driven spin diffusion~\cite{abragam,slichter,meier}) is based on the Bloch-Redfield theory. It assumes that the protons form a dense network of strongly coupled spins, possessing fast dynamics that quickly erases all correlations in the proton spin bath. Under this assumption, the action of the actual many-spin quantum system (the proton spin bath) is replaced by a classical time-varying random magnetic field, which leads to complete thermalization of the $^{13}$C spins. On the other hand, it is not unusual to see the results of the ssNMR experiments in stark disagreement with the predictions of the standard theory~\cite{NEVZOROV,thesis_koneke,BRUSCHWEILER,FELDMAN1998297,abanin}, and the origins of this disagreement are not properly understood. Our results demonstrate an important reason for possible failure of the standard spin diffusion theory.

It is natural to expect that the local noise at the C sites is determined by the nearby proton spins, so these protons will be most essential for driving the spin diffusion, while the protons located far away from the $^{13}$C sites, including the protons from other molecules, will have negligible effect. However, we show that this view is oversimplified because the "far-away" protons are critically needed for the collective ergodic dynamics of the noise created by the proton spin environment~\cite{polkovnikov,NEVZOROV,erg}. In particular, we show that, if the bath is not ergodic then the proton environment does not act like a noise, and thermalization is quenched. Thus, the "far-away" proton spins, in spite of contributing very little to the local noise, become as important as the nearby protons in thermalizing the $^{13}$C spins. 

\begin{figure}[t]
  \centering
  \includegraphics[width=0.7\columnwidth]{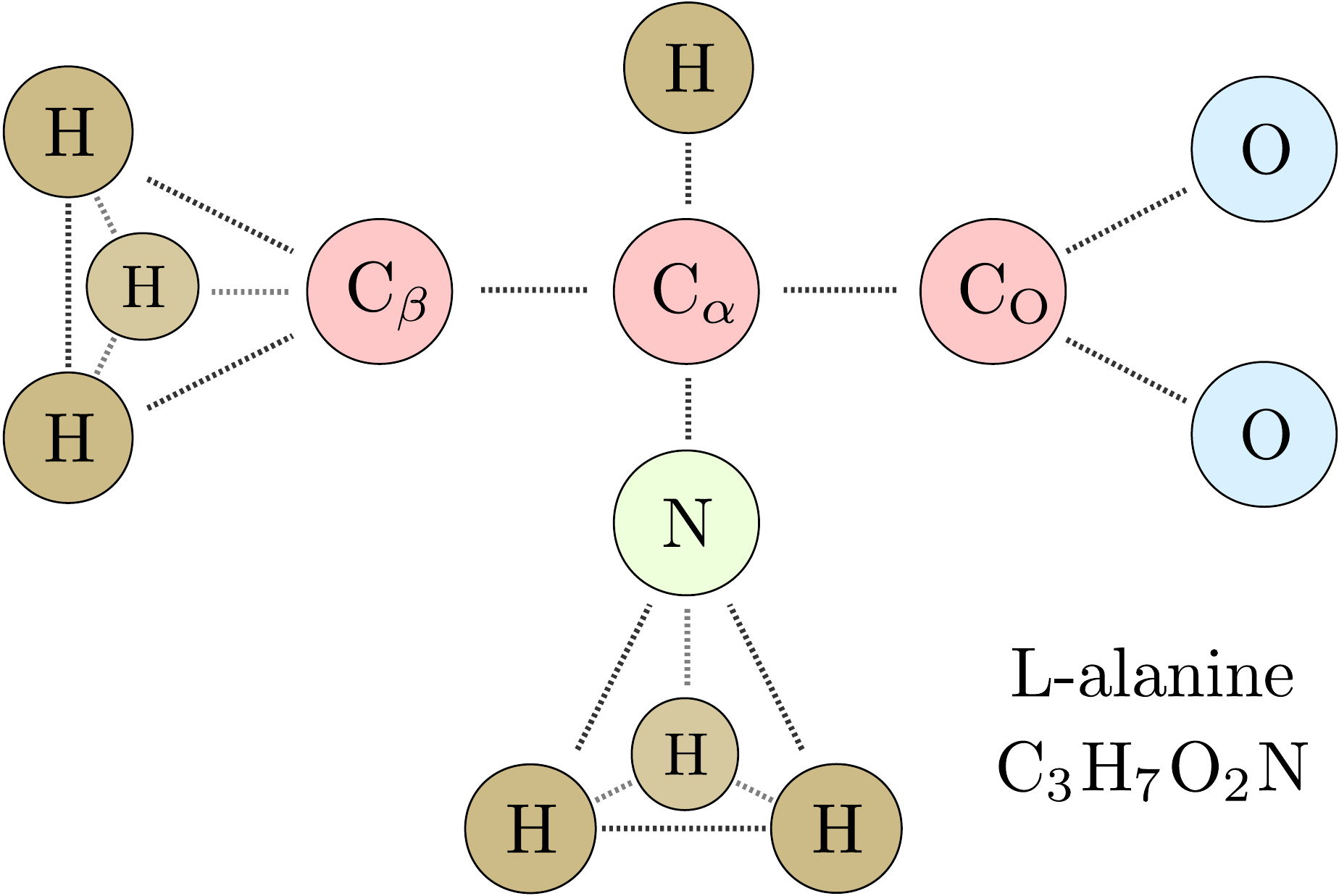}
  \caption{Schematic structure of the L-alanine molecule~\cite{simpson}. The three carbon nuclei are chemically nonequivalent and, therefore, experience different chemical shifts. The magnetic field at the position of each \isotope[13]{C} spin is governed by the \isotope[1]{H} spins (\isotope[15]{N} spin has a small magnetic moment and oxygen nuclei have no spin).}
 \label{fig_alanine}
\end{figure}

In order to make conclusive statements, we perform theoretical studies taking into account the conditions of real ssNMR experiments. 
We investigate thermalization in L-alanine (below we often drop the "L" prefix), starting with one polarized $\isotope[13]{C}$ spin in an otherwise unpolarized system (cf. Fig.~\ref{fig_alanine}), and watching the spreading of its polarization to the other $^{13}$C spins. In comparison with other systems typically studied in the context of thermalization, all parameters of this system are fixed (no random potential) and well known (no fitting), but, at the same time, the system is very complex, as is typical for ssNMR, exhibiting non-trivial thermalization and spin diffusion controlled by the collective many-body effects: it is 3-dimensional, includes long-range couplings, with the Hamiltonian modulated in time. 

%
%Besides fundamental importance, it is of practical value for improving the %ssNMR-based structural analysis and characterization.
%
%The direct dipolar interaction between the $^{13}$C spins includes the %Hamiltonian terms that drive the flip-flop transitions, but the large %difference in the chemical shifts between different C sites, and the resulting %difference in Zeeman energies, inhibits the direct transfer of polarization %between different $^{13}$C spins, so the thermalization within the %$\isotope[13]{C}$ subsystem is governed by the surrounding network of %$\isotope[1]{H}$ spins, including the protons belonging to different molecules %in the crystal. 
%

The rest of the article is organized as follows. In the next section, Sec.~\ref{sec_prob}, we provide a detailed description of the system and present qualitative discussion. In Sec.~\ref{sec_prel}, we describe in detail the numerical simulation methods. Results of our simulations are discussed in Sec.~\ref{sec_res_prot} and conclusions are given in Sec.~\ref{sec_conclude}.

\section{Description of the system} \label{sec_prob}

\subsection{Nuclear spins in alanine}

An L-alanine crystal has orthorhombic structure with the space group P$2_12_12_1$, and has four alanine molecules per unit cell. The unit cell dimensions are $a=6.0$~\AA, $b=12.3$~\AA\ and $c=5.8$~\AA. For the polycrystalline alanine powder, the isotropic average of individual single-crystallite alignments is considered~\cite{slichter,torchia,derbyshire}. A molecule of alanine has the chemical formula \isotope{C}$_3$\isotope{H}$_7$\isotope{O}$_2$\isotope{N}; its zwitterionic form structure~\cite{simpson} is illustrated in Fig.~\ref{fig_alanine}. Except for the oxygen nuclei which practically have no spin and are neglected in the following, all other nuclei \isotope[13]{C}, \isotope[15]{N} and \isotope[1]{H} have spins \textonehalf\ (since nitrogen spins are of little interest, for simplicity below we consider isotopically purified sample without \isotope[14]{N} spins). The three chemically nonequivalent carbon sites are commonly referred to as \mbox{$\isotope{C}_\text{O}$,} $\isotope{C}_\alpha$, and $\isotope{C}_\beta$. Among the nuclei, the protons have the largest magnetic moment, and correspondingly the largest gyromagnetic ratio
$\gamma_{\isotope{H}}\approx5.6$ rad/(s T); the gyromagnetic ratios of the other spins are significantly smaller:
\begin{equation} \label{eqn_gyro}
 \frac{\gamma_{\isotope{C}}}{\gamma_{\isotope{H}}}\approx 0.25,\quad  \frac{\gamma_{\isotope{N}}}{\gamma_{\isotope{H}}}\approx 0.1.
\end{equation}

\subsection{Hamiltonian of the system and characteristic energy scales}

We consider a polycrystalline alanine sample in a strong static magnetic field (quantizing field) $B_0$ along the $z$-axis, which sets the proton Larmor frequency at 400~MHz. The sample is subjected to the magic angle spinning (MAS), i.e.\ the sample is rotated with a frequency $\nu_r$ (below we take $\nu_r=10$~kHz) around the rotor axis which makes an angle of $54.7$\textdegree\ with the $z$-axis; MAS is often used in ssNMR to reduce the line width and increase spectral resolution~\cite{slichter,abragam}. 

The Hamiltonian of the system includes the single-spin Zeeman energies and the pairwise dipolar couplings. In the reference frame rotating with the Larmor frequency around the $z$-axis, the secular part of the Hamiltonian has the standard form~\cite{abragam,slichter}
\begin{equation} \label{eqn_ham}
 {\cal H}_\text{tot}(t)={\cal H}_\text{CS}(t)+{\cal H}_\text{DD}(t),
\end{equation}
where the bare Zeeman energies (the terms $\gamma_i B_0 S_{iz}$) of different spin species ($^{13}$C, $^{15}$N, and $^1$H) are eliminated by the rotating frame transformation. The time dependence of the Hamiltonian terms is caused by the MAS, and is periodic with the rotor period $T_r=1/\nu_r=0.1$~ms; correspondingly, all observables in the figures below are shown at times commensurate with $T_r$.

The term ${\cal H}_\text{CS}(t)$ describes the chemical shift, i.e.\ small deviation from the reference Larmor frequency of a given nuclear spin, caused by the finite electronic density near the nucleus: 
\begin{equation}
 {\cal H}_\text{CS}(t)=\sum_j \Delta\omega_{jz}(t)S_{jz}, \label{eqn_hamcs}
\end{equation}
where $\Delta\omega_{jz}=\delta_{zz,j}\gamma_j B_0$, and $\gamma_j$ is the gyromagnetic ratio of the $j$-th spin, and $\delta_{zz}$ is the $(z,z)$ component of the chemical shift tensor for the $j$-th site. The chemical shift is generally anisotropic; it is described by a symmetric rank-2 tensor~\cite{slichter}. The principal axes of this tensor are determined by the electronic density around the nucleus, and thus depend on the local crystalline symmetry. When the chemical shift tensor is transformed to the laboratory reference frame \cite{slichter,meier,grommek}, its $(z,z)$ component, and the resulting value of $\Delta\omega_{jz}$, depend on the orientation of the given crystallite with respect to the quantizing magnetic field ($z$-axis), and change in time because of the MAS \cite{slichter,meier,grommek}. In alanine, the chemical shift tensor for C$_\beta$ and C$_\alpha$ sites is moderatly anisotropic, its entries are of the order of 6--$12$~kHz, while for C$_\text{O}$ site it is strongly anisotropic, with the entries varying from +2 to -10~kHz \cite{naito}. The chemical shifts of the protons are isotropic, with the magnitude of order of 1--3~kHz \cite{HAFNER}. Here and below we follow the standard convention, setting $\hbar=1$, and expressing energies in rad/s or in Hz (1~Hz equivalent to 2$\pi$~rad/s).

The term ${\cal H}_\text{DD}(t)$ describes the dipole-dipole interaction
\begin{eqnarray}
 &&{\cal H}_\text{DD}(t)=\sum^\text{like}_{i,j}A_{ij}(t)\Big[S_{ix}S_{jx}+S_{iy}S_{jy}-2S_{iz}S_{jz}\Big] \nonumber \\
 &&\hspace{2cm} +\sum^\text{unlike}_{i,j}A_{ij}(t)\Big[-2S_{iz}S_{jz}\Big]. \label{eqn_hamdd}
\end{eqnarray}
The dipole-dipole couplings are calculated in a standard way using the coordinates of the atoms in the alanine crystal, i.e.\ 
$A_{ij}(t)=\gamma_i\gamma_j\hbar(1-3\cos^2{\theta_{ij}(t)})/r^3_{ij}$, where $\gamma_i$ and $\gamma_j$ are the gyromagnetic ratios of the $i$-th and $j$-th spin, $r_{ij}$ is the distance between them, and $\theta_{ij}$ is the polar angle of the vector connecting the $i$-th and $j$-th sites; note that this angle depends on the orientation of a given crystallite with respect to the $z$-axis, and changes in time in a periodic manner due to the MAS. The standard notations of "like" or "unlike" spins denote the spins belonging, respectively, to the same or different spin species (e.g.\ $^{13}$C and $^1$H), i.e.\ having the same or different gyromagnetic ratios.

The flip-flop transitions which cause the spin diffusion are mediated by the transverse dipolar coupling term $S_{ix}S_{jx}+S_{iy}S_{jy}=\frac{1}{2}[S^-_{i}S^+_j+S^+_{i}S^-_j]$, where $S^+_{i}$ and $S^-_i$ are rising and lowering operators. These terms are absent for the dipolar couplings between the unlike spins because the mismatch in Zeeman energies is large (hundreds of MHz), while the typical dipolar couplings rarely exceed few hundreds of kHz. 
In our specific system, within the strongly coupled groups, such as methyl (CH$_3$) and amino (NH$_3$) groups, the $^1$H-$^1$H and the $^{13}$C-$^1$H couplings are of the order of 30~kHz. The typical couplings between the protons belonging to different groups or different molecules are of the order of 5--10~kHz. 

The C-C couplings within the molecule are $\sim 2$~kHz for the C$_\alpha$-C$_\beta$ and C$_\text{O}$-C$_\beta$ pairs, i.e.\ noticeably smaller than the Zeeman energy differences caused by the chemical shifts. Thus, the interaction with the proton subsystem, which can absorb this energy mismatch, is critical for thermalization between the C spins.

At the characteristic temperatures of the ssNMR experiments, the protons within each methyl and ammonium group randomly interchange their positions at the timescale of the order of 1~ns \cite{torchia,thesis_long,beshah}. This is much faster than the characteristic times of the spin dynamics (microseconds to seconds). These fast jumps have to be accounted for by averaging ${\cal H}_\text{tot}(t)$ over the corresponding sites for all spins within each group.

\section{Numerical simulations of thermalization dynamics} \label{sec_prel}

We numerically simulate the dynamics of the many-spin system, which includes an alanine molecule and a number of protons around it; the total number of modeled spins was between $n_s=4$ (only C and N sites) and $n_s=24$ (one molecule and 13 protons around it). The surrounding protons are added individually for aliphatic sites, and in groups of three for methyl and ammonium groups. We directly solve the time-dependent Schr\"odinger equation with the Hamiltonian (\ref{eqn_ham}) for the many-spin wavefunction, represented as a $2^{n_s}$-dimensional vector, using the 4-th order Suzuki-Trotter expansion for the evolution operator~\cite{chebyshev,Suzuki1976}, with the timestep 2~$\mu$s.

The initial state of the system corresponds to the typical experiment, with one polarized $^{13}$C nuclear spin and the rest being unpolarized: different C sites are addressable due to their different chemical shifts~\cite{meier,thesis_koneke}, and can be individually polarized, manipulated, and measured. Since the typical temperatures of ssNMR correspond to the large-temperature limit for the nuclear spins, the normalized relevant part of the density matrix is \cite{slichter}
\begin{equation} \label{eqn_init}
 \rho_\text{init}=\Big[|\!\uparrow\rangle\langle\uparrow\!|\Big]_{\isotope{C}_\text{O}}\otimes\mathbb{I}_\text{rest},
\end{equation}
where the first term describes the \isotope{C}$_\text{O}$ spin and $\mathbb{I}_\text{rest}$, which is proportional to the unit matrix (so that $\text{Tr}[\mathbb{I}_\text{rest}]=1$), describes the rest of the system. 
In our simulations the density matrix $\mathbb{I}_\text{rest}$ was represented, with exponential precision, by a random wavefunction, i.e.\ by a $d$-dimensional  vector of complex numbers, with the entries drawn randomly from the uniform distribution on a $(d-1)$-dimensional complex unit sphere~\cite{goldstein,gemmer}, where $d=2^{(n_s-1)}$ is the dimensionality of the Hilbert subspace corresponding to the unpolarized rest of the system.

To model the standard polycrystalline powder sample, we averaged the spin polarizations over a large number (in most cases, over 200) of randomly oriented crystallites. The quantity of interest for us is the time-dependent polarization $P_z(t)\equiv {\rm Tr}\left[2S_z\rho(t)\right]$ on different C sites. Since the flip-flops occur only between the $^{13}$C spins, the total polarization of the $^{13}$C subsystem remains constant. In the case of complete thermalization, at long times we have
\begin{equation} \label{eqn_thermal}
 P_{z}[\isotope{C}_{\text{O}}](t)\approx P_{z}[\isotope{C}_\beta](t)\approx P_{z}[\isotope{C}_\alpha](t)\approx\frac{1}{3}.
\end{equation}

A few notes are in order. First, our test simulations have shown that the results obtained with two alanine molecules and several protons around them are practically the same as the results obtained with a single molecule and protons around it. The spin diffusion between the C sites of different molecules is small, and the dynamics of all $^{13}$C spins in the sample can be reproduced by modeling a single molecule. Second, we also simulated the spin diffusion with other initially polarized $^{13}$C spins (not shown here); the simulations show that our conclusions about the spin diffusion rates and emergence/quenching of thermalization do not depend on the choice of the initially polarized C site.

\begin{figure}[t]
  \centering
  \includegraphics[width=0.9\columnwidth]{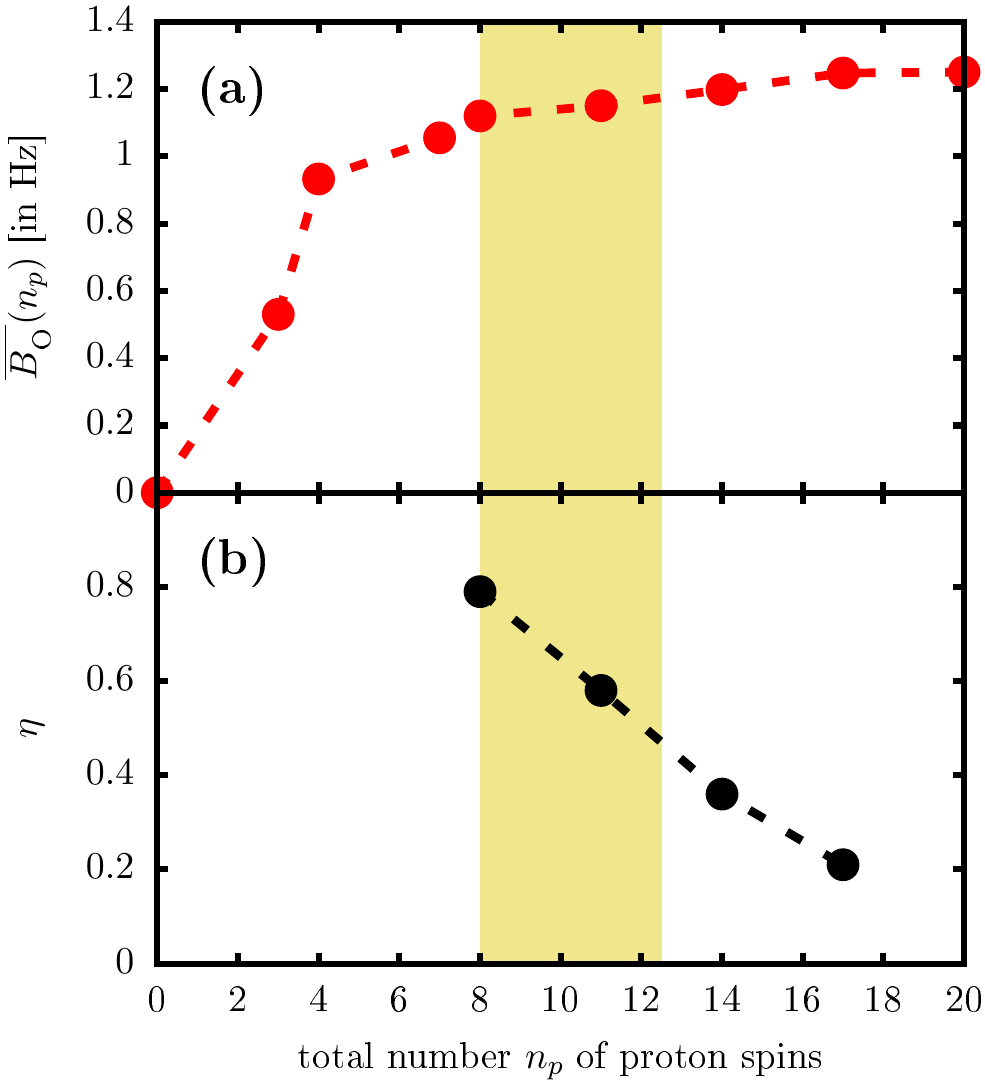}
  \caption{\textbf{(a)} Effective magnetic field $\overline{B_\text{O}}(n_p)$ created at the $\isotope{C}_\text{O}$ site by the proton spins. The first seven proton spins belong to the same molecule as $\isotope{C}_\text{O}$, the further spins from neighboring molecules were included according to their distance to $\isotope{C}_\text{O}$, cf. Eq.~\eqref{eqn_eff_field}. The effective magnetic field $\overline{B_\text{O}}(n_p)$ is mainly governed by the proton spins from the same molecule, whereas the spins from neighboring molecules produce only a rather small correction to $\overline{B_\text{O}}(n_p)$. \textbf{(b)} The chaoticity parameter $\eta$ introduced in Eq.~\eqref{eqn_eta} significantly decreases with increasing $n_p$. In \textbf{(a)} and \textbf{(b)}, dashed lines connect points to guide the eye.}
 \label{fig_loc_field}
\end{figure}

\section{spin diffusion in alanine: numerical results} \label{sec_res_prot}

\subsection{Magnitude of local fields}

% [FIX THE AXES AND NOTATIONS IN FIG 2 ???]

The spin diffusion is controlled by the (quasi-)random magnetic field created by the surrounding proton spins at the carbon sites, e.g.\ at the C$_\text{O}$ site:
\begin{equation}
B_O(t)=-2\sum_{j\in {\cal P}} A_{j,O}(t) S_{jz},
\end{equation}
where the summation is over the set $\cal P$ of proton sites, and $A_{j,O}(t)$ is  the parameter of the dipolar coupling between the C$_\text{O}$ spin and the $j$-th proton (including the time modulation by MAS), see Eq.~\ref{eqn_hamdd}. It is instructive to calculate the magnitude of this field as a function of the number $n_p$ of protons included in the modeled system. The average of this field is zero for unpolarized protons, while the second moment
\begin{equation} 
\overline{B_O}^2 \equiv \frac{1}{T_r} \int_0^{T_r} dt \int_{\Omega} d\mu_R {\rm Tr}\left[\rho(t) B_O^2(t)\right]
\end{equation}
is finite, and includes quantum-mechanical averaging, time averaging over the rotor period $T_r$, and averaging over the crystallite orientation (i.e. the integral over the group $\Omega$ of all 3D rotations $R$ of the crystallite, weighted with the ter Haar measure $d\mu_R$). The dispersion of the random field $\overline{B_O}(n_p)$ can be expressed as
\begin{equation}
\label{eqn_eff_field}
\overline{B_O} (n_p) = \sqrt{\sum_{j=1}^{n_p} \overline{A}_{j,O}^2}, \quad
\overline{A}_{j,O}^2=\frac{1}{T_r}\int_0^{T_r} dt \langle A_{j,O}^2 (t)\rangle_R,
\end{equation}
and can be directly calculated from the known crystal and molecular structure, 
where $\langle\dots\rangle_R$ denotes average over the crystallite orientations.
Since the dipolar coupling decays with distance $r$ as $1/r^3$, and its square as $1/r^6$, the variance for an infinite (i.e.\ macroscopic-sized) sample $\overline{B_O}(\infty)$ is well defined, and is determined almost completely by only several (7--10) nearby protons. This dependence is shown in Fig.~\ref{fig_loc_field}, where 
$\overline{B_O} (n_p)$ is plotted for the C$_\text{O}$ site and $n_p$ increases as the protons are added one by one according to their distance from C$_\text{O}$. 

\begin{figure}[b]
  \centering
   \includegraphics[width=0.9\columnwidth]{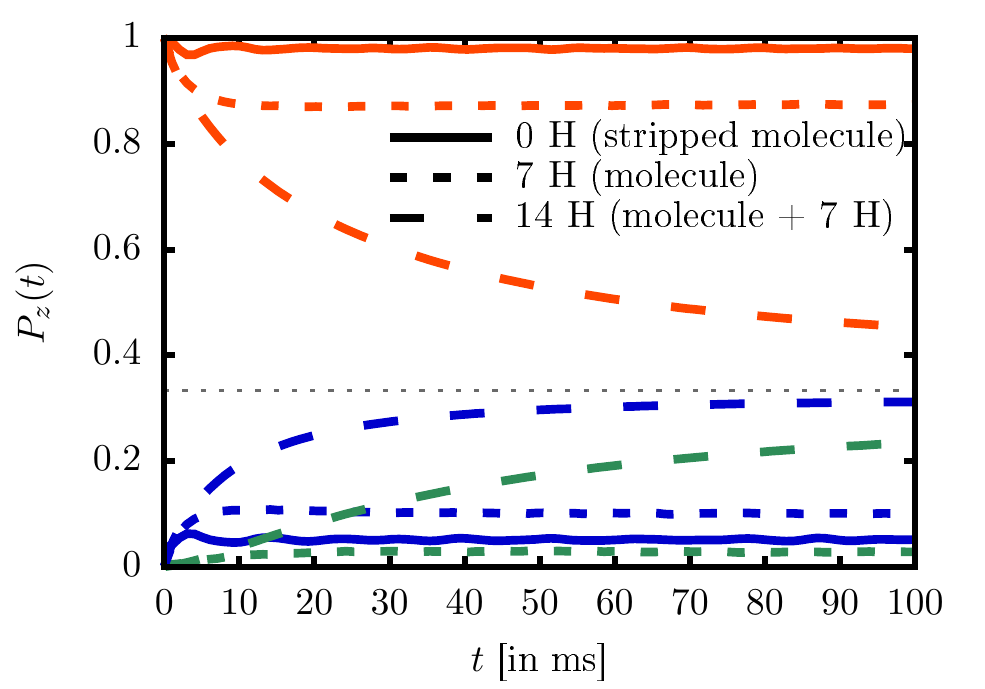}
  \caption{Influence of the surrounding proton bath on spin diffusion in alanine. The time evolution of $P_{z} (t)$ for individual carbon spins is shown for a different number of proton spins considered in the simulation. The color coding of the carbon spins is: $\isotope{C}_\text{O}$ - red, $\isotope{C}_\alpha$ - blue, and $\isotope{C}_\beta$ - green. The system simulated consists of three carbon spins and one nitrogen spin (solid line), one bare molecule (dashed line), a single molecule with seven nearest proton spins from neighboring molecules (fine dashed line). The horizontal dotted gray line indicates the equilibrium value 1/3.}
 \label{fig_prot_few}
\end{figure}

\begin{figure}[t]
  \centering
   \includegraphics[width=0.9\columnwidth]{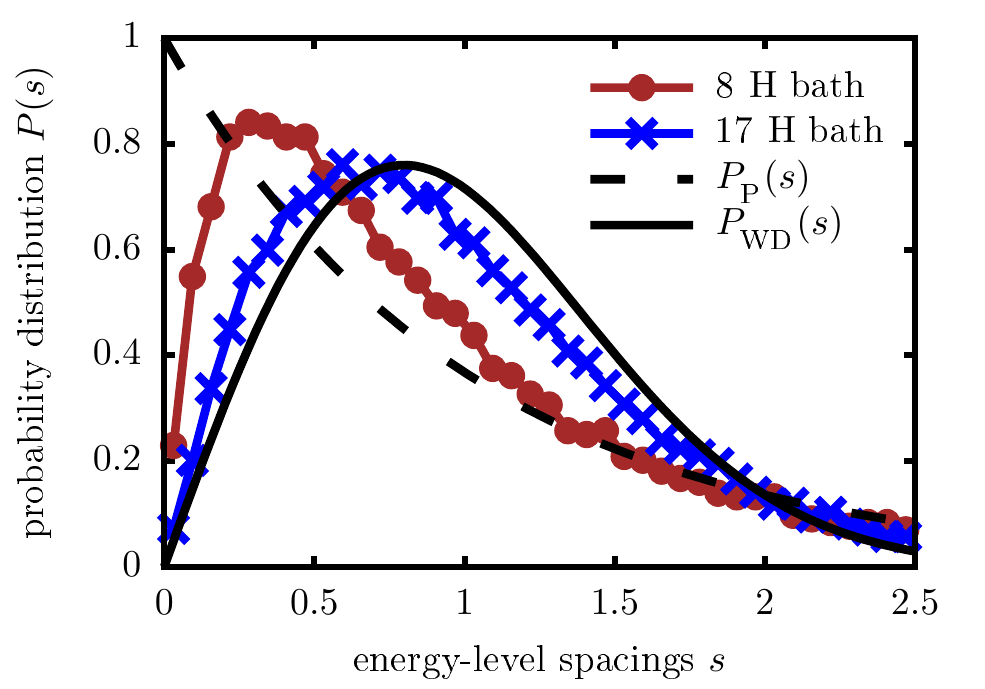}
  \caption{Transition from integrability to nonintegrability in the proton bath with increasing bath size $n_p$. The normalized probability distribution $P(s)$ of energy-level spacing $s$ of the Hamiltonian describing the proton bath approaches the Wigner-Dyson distribution $P_\text{WD}(s)$ as we increase $n_p$. The distributions $P(s)$ are averaged over crystallite alignments (number of alignments - $n_p=8$: 1000, $n_p=17$: 8). The results are obtained for one irreducible block of the Hamiltonian taken at $t=0$. We do not average the Hamiltonian over the proton sites in each individual group to exclude additional effective constants of motion. Lines connect point to guide the eye.}
 \label{fig_spacing}
\end{figure}

\subsection{Spin diffusion between C sites}

While the "far away" protons add very little to the local field at the C sites, they play crucial role in the spin diffusion in the $^{13}$C subsystem, as seen from Fig.~\ref{fig_prot_few}. We start from simulating thermalization in a single molecule with all protons removed, and observe an almost static $P_{z}(t)$ (Fig.~\ref{fig_prot_few}, solid line), as expected: without protons the difference in the chemical shifts between the C sites is too large for spin  diffusion to happen. Next, we add protons, and simulate a single molecule of alanine (dashed line), which contains seven protons. The second moment of the local field, given by Eq.~\ref{eqn_eff_field} above, for $n_p=7$ reaches 80\% of its maximum value. Initially, the polarization $P_z(t)$ diffuses from C$_\text{O}$ to other sites, but after $\approx 5$~ms it saturates at the values which are very far from thermalization. With further increasing the number of proton spins, the saturation value of $P_{z}$ for $\isotope{C}_\text{O}$ decreases, but still does not reach its thermalized value 1/3. Only when we include 7 nearest protons from other molecules, we see a onset of thermalization in the $^{13}$C subsystem, although the magnitude $\overline{B_O}$ of the random field changes very little in this range of $n_p$. 
After that, including more and more protons in the simulations does not change the behavior of $P_z(t)$ much. These results clearly demonstrate our main point: the "far away" proton spins significantly affect spin-diffusion dynamics almost without affecting the local field $\overline{B_\text{O}}(n_p)$.

In order to identify the key factor controlling the thermalization, we investigate an onset of chaos in the dynamics of the proton bath as the number of protons changes from $n_p=7$ (single alanine molecule, no thermalization) to $n_p=14$ (a molecule with 7 additional protons, thermalization). As a signature of chaos, we use the statistics $P(s)$ of the nearest level spacings $s$ \cite{guhr,mehta,shepelyansky_eta,santos_vector,friedrich}, which is a standard spectral measure of non-integrability in quantum systems. As seen in Fig.~\ref{fig_spacing}, for $n_p=8$ protons the shape of $P(s)$ is close to the Poisson distribution $P_\text{P}(s)=\exp(-s)$, characteristic for integrable quantum system. As $n_p$ increases, the shape of $P(s)$ gradually changes, and for $n_p=17$ almost coincides with the orthogonal Wigner-Dyson distribution $P_\text{WD}(s)=s\frac{\pi}{2}\exp(-s^2\frac{\pi}{4})$, which is a hallmark of the quantum chaos in the system.
The closeness of the distribution $P(s)$ to the chaotic or integrable case can be quantified with the parameter 
\begin{equation} \label{eqn_eta}
 \eta=\frac{\bigintsss_0^{s_0}\Big[P(s)-P_\text{WD}(s)\Big]ds}{\bigintsss_0^{s_0}\Big[P_\text{P}(s)-P_\text{WD}(s)\Big]ds},
\end{equation}
where $s_0$ is defined as the smaller value satisfying $P_\text{P}(s_0)=P_\text{WD}(s_0)$~\cite{shepelyansky_eta,lages}; so that $\eta=1$ corresponds to a purely Poisson distribution, while $\eta=0$ corresponds to the Wigner-Dyson statistics. Fig.~\ref{fig_loc_field}(b) shows that the parameter $\eta$ monotonically decreases from 0.8 to 0.4 in the same region $n_p$ where the onset of thermalization within the C subsystem is observed.

\begin{figure}[b]
  \centering
  \includegraphics[width=0.9\columnwidth]{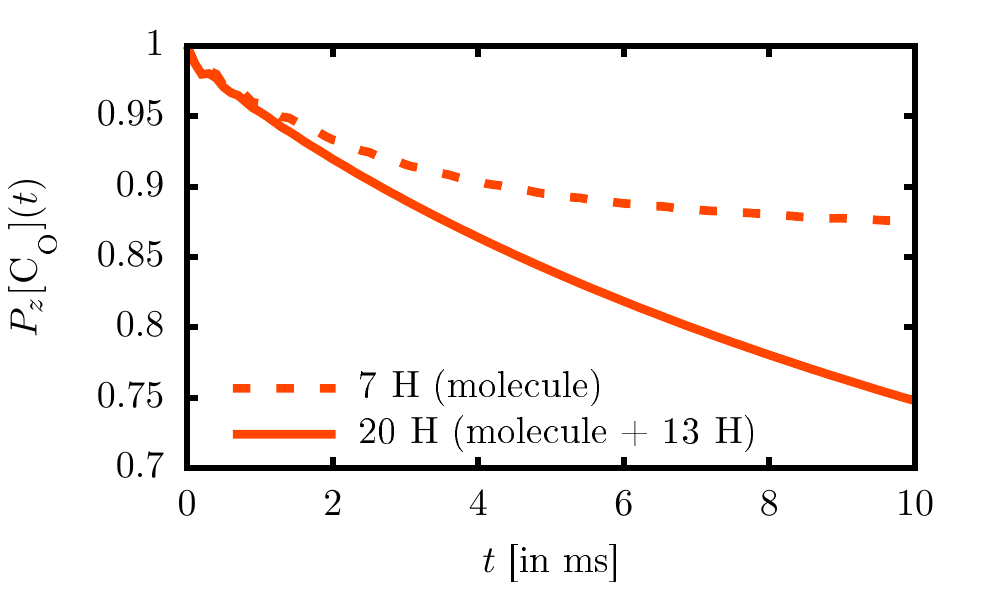}
  \caption{Initial \mbox{vs.} long-time behavior of the time evolution of $z$-polarization of $\isotope{C}_\text{O}$ spin for different sizes of the surrounding proton bath as indicated in the legend. In contrast to longer time, the initial spin diffusion rate ($t\leq2$ ms) does not depend on the size of the proton bath. For a proton bath consisting of less than 14 protons, $P_{z}[\isotope{C}_{\text{O}}](t)$ saturates at non-equilibrium values.}
 \label{fig_prot_init}
\end{figure}

It is also instructive to look closer at the initial change of the spin polarization $P_z(t)$ at the C$_\text{O}$ site, see Fig.~\ref{fig_prot_init}. For $n_p\ge 7$, we observe that the initial rate of spin diffusion does not change much as the number of protons in the simulated system increases. This means that the initial diffusion rate stays approximately the same as long as the local-field dispersion $\overline{B_O}(n_p)$ does not vary much. This is exactly what one would qualitatively expect from the Bloch-Redfield theory, which predicts that the rate $\Gamma$ of the polarization transfer between two C sites is \cite{meier,thesis_koneke}
\begin{equation}
\label{eq_sdrate}
\Gamma\propto A_{12}^2 F_{ZQ}(\delta),
\end{equation}
where $A_{12}$ is the dipolar coupling between the two C sites, $\delta$ is the difference in the chemical shifts, and $F(\delta)=\int f_1(\omega) f_2(\delta-\omega) d\omega$ is the overlap of the zero-quantum NMR lines of the two $^{13}$C spins, given by the convolution of their zero-quantum lineshapes $f_1(\omega)$ and $f_2(\omega)$~\cite{meier}. The zero-quantum lines overlap $F_{ZQ}$ is controlled primarily by the local random fields, so the rate $\Gamma$ should not change much if $\overline{B_O}$ does not change.

\begin{figure}[t]
  \centering
   \includegraphics[width=0.9\columnwidth]{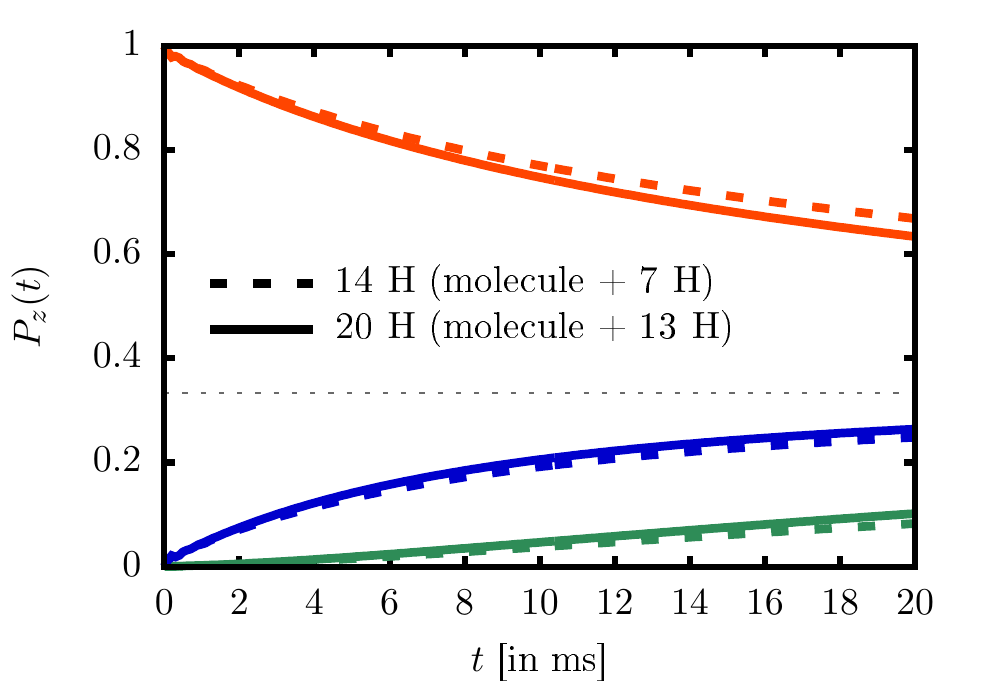}
  \caption{Small modifications of spin-diffusion behavior in alanine for larger spin bath. The time evolution of $P_{z} (t)$ of individual carbon spins is shown for a different number of proton spins considered in the simulation. The color coding of the carbon spins is: $\isotope{C}_\text{O}$ - red, $\isotope{C}_\alpha$ - blue, and $\isotope{C}_\beta$ - green. The system simulated consists of a single molecule with 13 (solid line) or 7 (dashed line) additional nearest proton spins from neighboring molecules. The horizontal dotted gray line indicates the equilibrium value 1/3.}
 \label{fig_prot_large}
\end{figure}

This standard logics breaks down at later times, when $P_z(t)$ saturates. For small number of protons, in spite of the same initial depolarization rate, the curve $P_z(t)$ saturates quickly and far from the thermalized value, since the overall dynamics of the proton bath is far from chaotic, and the local fields are far from random. In this regime, adding even a single extra proton changes the behavior of $P_z(t)$ at $t\ge 4$~ms considerably, and noticeably lowers the saturation value. 
For larger proton baths, once chaotic behavior emerges, thermalization sets in, and $P_z(t)$ saturates at the value 1/3. After that, adding more proton spins from neighboring molecules does not change $P_z(t)$ significantly, as seen in Fig.~\ref{fig_prot_large}: even though $n_p$ increases from 14 to 20, the changes in the $P_z(t)$ are small even at very long times.

% [SHOW SIMULATIONS UNTIL 300 MS???]

\section{Conclusions} \label{sec_conclude}

We conclude that the nearby protons and the far-away proton spins, while both essential for the spin diffusion process, play very different roles. The nearby protons govern the dispersion of the local fields, and thus determine the  short-time behavior \mbox{$0\leq t\leq4$ ms} of $P_{z}(t)$. The long-time dynamics and thermalization between the C spins is governed by the far-away protons, which do not directly affect the local fields, but control the collective many-body dynamics of the proton bath. They ensure emergence of quantum chaos among the proton spins, and, as the spectral statistics $P(s)$ shows, it is the emerging chaos in the proton bath that triggers thermalization among the $^{13}$C spins. In this way, the far-away protons ensure the very existence of the strongly coupled proton spin bath, which is assumed in the Bloch-Redfield theories of spin diffusion.

The experimental ssNMR tests of this idea are possible by using the deuterated alanine samples, since deutrons have a noticeably smaller magnetic moment, and the dipolar coupling between deutron spins is about 10 times smaller than the coupling between nuclear spins. However, the quadrupolar coupling, and the related fast spin-lattice relaxation characteristic for the deutrons (which have spin 1, in contrast with the proton spin 1/2), may also modify the spin diffusion between the $^{13}$C spins; these effects should be taken into account in the future simulations.

\section*{Acknowledgments}
We thank M. Hong and K. Schmidt-Rohr for helpful discussions. This work has been made possible by the Foundation for Fundamental Research on Matter (FOM), which is part of the Netherlands Organisation for Scientific Research (NWO).

\end{document}